\documentclass{article}

\usepackage{vendor/iclr2027/iclr2027_conference,times}
\usepackage{amsmath,amssymb,booktabs,graphicx,microtype,needspace,placeins}
\usepackage{algorithm,algpseudocode}
\usepackage[hidelinks]{hyperref}
\usepackage{url}

\newcommand{\SYS}{\textsc{RR-Evict}}

\title{\SYS{}: Fine-Grained Prefix Cache Eviction beyond LRU for Agentic LLM Serving}

\author{
Zaifeng Pan$^{1}$\thanks{Correspondence to: Zaifeng Pan $<$zapan@ucsd.edu$>$.} \enspace
Chris Wu$^{1}$ \enspace Zhengding Hu$^{1}$ \enspace
Xinwei Qiang$^{1}$ \enspace Zhongkai Yu$^{1}$ \enspace Yufei Ding$^{1}$ \\
$^{1}$University of California San Diego
}

\iclrfinalcopy

\begin{document}
\maketitle
\lhead{Preprint. Under review.}

\begin{abstract}
LLM-based agents execute long-horizon tasks through repeated model calls interleaved with tool execution and user interaction. As each call extends the history accumulated in previous turns, prefix caching avoids repeated prefill of the agent's entire context. However, the aggregate cache footprint grows with context length and concurrency, forcing serving systems to reclaim cached KV tensors. We identify \emph{recency synchronization}: accesses to an agent's cached history refresh its cache nodes together, causing least recently used (LRU) eviction to concentrate on a few agents' private histories. When a fully evicted agent returns, it must recompute nearly its entire accumulated context, producing a large time-to-first-token (TTFT) outlier even if most other requests retain substantial cache reuse. We present \SYS{}, a fine-grained prefix-cache eviction strategy that distributes reclamation across idle agent trajectories. \SYS{} visits agents in round-robin order and evicts a tail chunk from each, preserving reusable prefixes for more agents when capacity remains for idle state. Returning agents can reuse these partial histories and recompute smaller missing suffixes. The policy requires no prediction of future arrivals or tool latency. We implement \SYS{} in SGLang and evaluate conversational and coding-agent workloads under colocated and prefill--decode-disaggregated serving. Compared with LRU, \SYS{} reduces P99 TTFT by up to 75.4\% and P99 uncached prompt tokens by up to 65.7\%.
\end{abstract}

\section{Introduction}
\label{sec:introduction}

Large language model (LLM) agents~\citep{yao2023react,shinn2023reflexion,yang2024swe,wang2025openhands} are increasingly used to execute long-horizon tasks that span many rounds of model inference, tool invocations, and user interactions. Unlike conventional stateless requests, each new agent request typically extends the history accumulated in previous rounds. As a result, the input context grows over the agent's lifetime, potentially reaching hundreds of thousands or even millions of tokens~\citep{packer2023memgpt,anthropic_context_engineering,context_for_deepagents}.

Prefix caching~\citep{sglang,gim2024prompt,gao2024cost,yu2025stateful,pan2025kvflow} is essential for serving such workloads efficiently. After an agent completes an LLM invocation and starts executing a tool or waiting for user feedback, the serving system can retain the key-value (KV) cache associated with its context. When the agent issues its next request, the system reuses the cached KV tensors for the shared prefix and computes the uncached suffix. If the previous history remains cached, this suffix contains only the newly appended tokens. Prefix caching therefore avoids repeatedly prefilling the agent's entire history and reduces time to first token (TTFT).

However, retaining prefix caches for long-horizon agents introduces significant GPU memory pressure~\citep{qin2025mooncake,liu2025lmcache,xie2026strata}. The cache footprint of each agent grows with its context, while the aggregate footprint increases further with the number of concurrent agents. When the total demand exceeds the available capacity, existing LLM serving systems typically reclaim cached state using least recently used (LRU) eviction~\citep{sglang,vllm_prefix_cache}.
LRU orders individual cache entries by their last-access times and is effective when different entries exhibit heterogeneous reuse patterns. This is often the case in conventional LLM serving workloads. For example, a system prompt shared by many requests is accessed frequently, whereas request-specific suffixes may be accessed only once. Their last-access times therefore provide a useful signal for distinguishing broadly reusable prefixes from less valuable suffixes. By evicting the least recently accessed entries, LRU can preserve frequently reused prefixes while incrementally reclaiming less reusable parts of the cache.

Long-horizon agent workloads change this access pattern. Each new request from an agent extends its previous context and therefore matches the agent's entire surviving cached history. Prefix matching refreshes all cache nodes along this path at approximately the same time, regardless of when those nodes were originally created or how much recomputation each represents. Moreover, different agents typically share only a short common prefix, while most of their accumulated histories remain private. Consequently, the private cache nodes of each agent form a large group with nearly identical recency. We refer to this phenomenon as \emph{recency synchronization}.

Under recency synchronization, nominally node-level LRU can behave like coarse-grained, agent-level eviction. When the system needs to reclaim more memory than a small number of cache nodes provide, LRU repeatedly selects nodes belonging to the oldest idle agent. It therefore tends to remove most or all of that agent's cached context before moving on to another agent. Although reclamation still operates on individual nodes, cache loss becomes concentrated on entire agent histories.

Such concentrated eviction is particularly harmful to tail latency. When a fully evicted agent becomes active again, it receives little prefix-cache reuse and must recompute nearly its entire accumulated history. Because the recomputation cost grows with context length, a small subset of agents can experience severe TTFT spikes even when most other requests retain substantial reuse. Therefore, optimizing only aggregate cache reuse is insufficient for long-horizon agent serving. The system must also consider how cache loss is distributed across agents.

To address this problem, we present \SYS{}, a fine-grained prefix-cache eviction strategy for long-horizon agent workloads. Instead of reclaiming one agent's entire context at a time, \SYS{} distributes eviction across idle agents in a round-robin manner. Whenever memory reclamation is required, it evicts a tail chunk from one agent and then advances to the next, continuing cyclically until sufficient memory has been released. Evicting from the tail preserves a valid, contiguous cached prefix, while round-robin selection distributes the reclamation demand across multiple histories.

This design replaces concentrated cache loss with gradual degradation. When capacity remains for idle state, \SYS{} retains partial histories for more agents under the same reclamation demand. Returning agents can reuse their remaining prefixes and recompute only the missing suffixes instead of prefilling their entire histories. By balancing cache loss across agents, \SYS{} reduces both the frequency and magnitude of cold-prefill events, thereby improving tail TTFT without requiring knowledge of future agent execution.

We implement \SYS{} in SGLang~\citep{sglang} and evaluate Qwen3~\citep{yang2025qwen3} on NVIDIA H100 GPUs using conversational and coding-agent workloads. Across colocated and prefill--decode (PD) disaggregated configurations, \SYS{} reduces P99 TTFT by up to 75.4\% and P99 uncached prompt tokens by up to 65.7\% compared with LRU. Comparisons with completion-aware LRU isolate the additional benefit of distributing eviction across live histories. Request-level distributions and a trace-derived memory-demand timeline further connect these gains to the policy's operating regime.

In summary, this paper makes the following contributions:
\begin{itemize}
    \item We identify \emph{recency synchronization} in long-horizon agent workloads and explain how it makes node-level LRU concentrate cache loss on a few trajectories, producing large tail-TTFT outliers.
    \item We propose \SYS{}, a fine-grained eviction strategy that reclaims tail chunks in round-robin order. We characterize when preserving partial prefixes for more agents can reduce tail recomputation.
    \item We implement \SYS{} in SGLang and evaluate conversational and coding agent benchmarks, showing improvements over both LRU and completion-aware LRU across the tested configurations.
\end{itemize}

\section{Background}
\label{sec:background}

\begin{figure}
    \centering
    \includegraphics[width=0.8\linewidth]{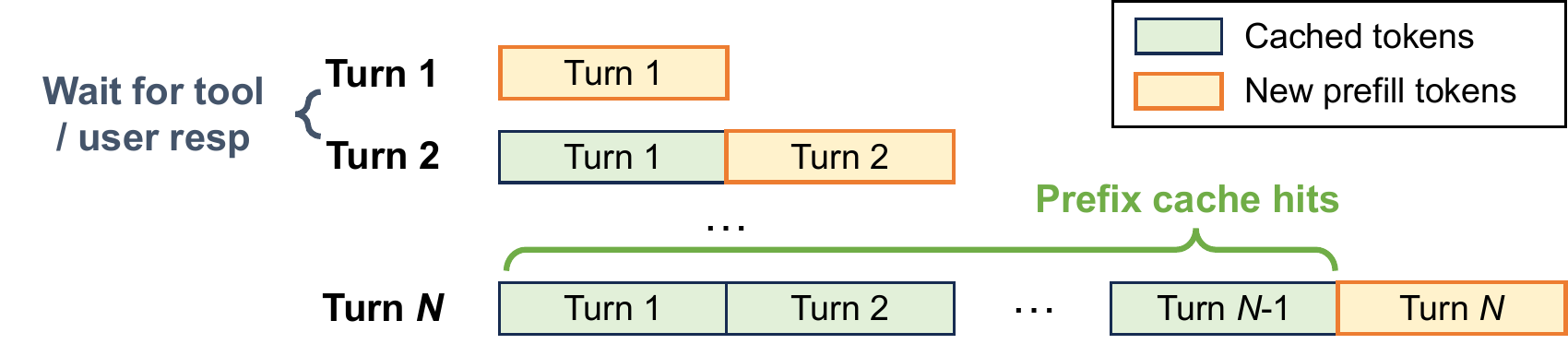}
    \caption{Prefix caching across agent turns. The KV tensors of the accumulated history are reused, and only newly appended tokens require prefill computation.}
    \label{fig:prefix-caching}
\end{figure}

\paragraph{Multi-turn agent workloads.} LLM agents solve long-horizon tasks through repeated interactions with users and external environments~\citep{yao2023react,shinn2023reflexion,yang2024swe,wang2025openhands,hu2026tritondft}. Each LLM invocation, which we call an \emph{agent turn}, reasons over the current context and produces a response or tool call.
As Figure~\ref{fig:prefix-caching} shows,
a turn may be followed by tool execution or a wait for user feedback, during which the agent issues no LLM request. The returned observation is appended to the next request together with the preceding history. Agent workloads are therefore both stateful and intermittent: requests from the same agent share an evolving context but are separated by variable external delays~\citep{anthropic_agent}. As the agent proceeds, its context grows while retaining most tokens from the previous turn.


\paragraph{Prefix caching for multi-turn agents.} LLM inference consists of a prefill phase followed by autoregressive decoding~\citep{yu2022orca}. During prefill, the model processes the input tokens and materializes their key-value (KV) tensors in every attention layer. Prefix caching retains these tensors across requests and reuses them when a later request begins with the same token sequence~\citep{sglang,gim2024prompt,gao2024cost,yu2025stateful}. As Figure~\ref{fig:prefix-caching} illustrates, Turn~1 computes its entire input, whereas Turn~$N$ can reuse the KV tensors for Turns~1 through~$N-1$ and prefill only the newly appended tokens. A longer prefix-cache hit avoids more recomputation and reduces time-to-first-token (TTFT).

Modern serving systems manage reusable KV tensors in a shared cache pool~\citep{vllm,sglang,vllm_prefix_cache}. Systems such as SGLang organize cached sequences as a radix tree, allowing requests to reuse the KV blocks along their longest matching prefix. In agent workloads, short prefixes such as system instructions may be shared across agents, while the longer interaction histories are typically private to each agent. As these histories and the number of concurrent agents grow, their aggregate demand can exceed the pool's GPU memory capacity. The eviction policy then determines how much history remains reusable when each agent resumes and, consequently, its next-turn prefill latency.


\section{\SYS{}: Fine-Grained Prefix-Cache Eviction}
\label{sec:method}

In this section, we first examine why conventional LRU eviction is inefficient for long-horizon agent workloads and then present \SYS{}, which distributes cache reclamation across agents at fine granularity. Finally, we characterize the operating regime in which round-robin eviction improves tail latency. Figure~\ref{fig:lru-rr} contrasts the two policies.

\subsection{Inefficiency of LRU Eviction}
\label{sec:lru-limitation}

Serving systems index reusable KV tensors at different granularities. SGLang uses radix-tree nodes representing variable-length context spans, whereas vLLM manages fixed-size KV blocks~\citep{sglang,vllm_prefix_cache}. We use \emph{cache node} to denote an entry in the prefix index. The following description focuses on SGLang's tree representation.

\begin{figure}[t]
    \centering
    \includegraphics[width=0.9\linewidth]{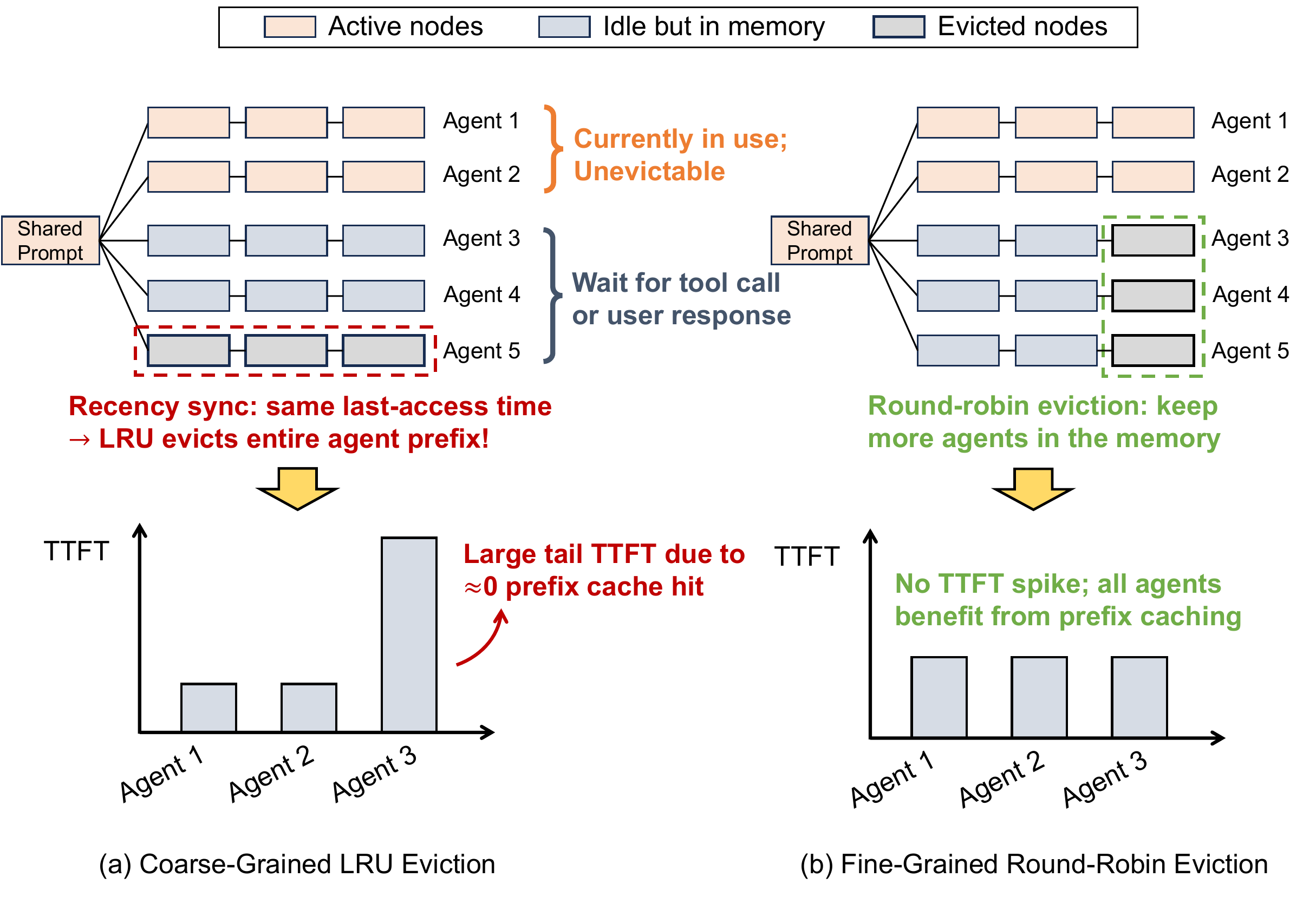}
    \caption{Schematic comparison of LRU and \SYS{}. (a) Recency synchronization concentrates successive LRU evictions on one idle agent, causing a large prefill when it returns. (b) Round-robin tail eviction preserves partial prefixes for more agents. The TTFT bars illustrate the mechanism rather than measured latencies.}
    \label{fig:lru-rr}
\end{figure}

As shown in Figure~\ref{fig:lru-rr}(a), nodes referenced by executing requests are active and cannot be reclaimed, whereas the cached trajectories of agents waiting for tool outputs or user responses are idle and eligible for eviction. To preserve a valid cache tree, eviction begins at idle leaf nodes and progressively exposes their parents as new leaves. Conventional LRU inserts eligible leaves into a heap ordered by their last-access time. This node-level policy is effective when nodes have sufficiently diverse access times because it can incrementally reclaim the least recently used data.

Agent workloads reduce this diversity. When an agent starts a turn, prefix matching accesses every node along its surviving cached prefix, updating those nodes at approximately the same time. After the turn completes, its private path becomes eligible for eviction with identical or nearly identical recency. Because agents typically share only a short system prompt while keeping most of their histories private, each agent contributes a large group of private nodes whose recency is synchronized. This is the \emph{recency synchronization} described in Section~\ref{sec:introduction}.

During reclamation, the synchronized nodes of the oldest idle agent remain at or near the top of the LRU heap as their descendants are removed and their parents become leaves. Successive evictions therefore concentrate on that agent until most or all of its private prefix is removed, effectively turning the agent into the unit of eviction. When the agent resumes, it can reuse little beyond the short shared prompt and must prefill nearly its entire history. As illustrated at the bottom of Figure~\ref{fig:lru-rr}(a), this concentrated cache loss produces a large TTFT spike for the evicted agent even when the aggregate cache hit rate remains high.

\subsection{\SYS{}: Round-Robin Tail Eviction}
\label{sec:rr-eviction}

\SYS{} retains the existing cache-tree traversal and active-node filtering but changes how eviction candidates are ordered. When memory must be reclaimed, the system scans the leaf nodes and skips those currently used by active requests. Instead of inserting the remaining leaves into a global heap ordered by LRU priority, \SYS{} groups them by their agent trajectories and visits these groups in round-robin order. It reclaims one \emph{tail chunk} from an agent before moving to the next. A tail chunk is the backend-specific amount reclaimed in one step and may be an entire leaf node or only a suffix within that node. A partially reclaimed leaf remains the trajectory's next candidate. Removing it completely exposes its parent, which can be selected only if it is also eligible for eviction. The process continues until the requested amount of memory has been released.

Round-robin selection converts concentrated cache loss into gradual degradation across agents. Under the same reclamation demand, several agents can retain partial histories rather than one agent losing its complete history. As illustrated at the bottom of Figure~\ref{fig:lru-rr}(b), returning agents can reuse their surviving prefixes and prefill only the missing suffixes. This reduces concentrated cold-prefill events when capacity remains for idle state. \SYS{} requires no prediction of future arrivals or tool latency. It uses only the current active/idle state and cache-tree metadata.

Algorithm~\ref{alg:rr-eviction} summarizes the round-robin policy. At the start of reclamation, eligible idle trajectories enter a single queue. After evicting one tail chunk, the algorithm moves the trajectory to the back of the queue if it still has an evictable suffix. Eligibility is rechecked before every eviction because an agent may resume concurrently. The backend-specific \textsc{SelectTailChunk} operation chooses either a complete leaf or a suffix of one.

\begin{algorithm}[H]
    \caption{Round-Robin Tail Eviction}
    \label{alg:rr-eviction}
    \begin{algorithmic}[1]
        \Require Idle agent trajectories $\mathcal{A}$, memory to reclaim $M$
        \State $Q \gets$ queue of trajectories in $\mathcal{A}$ with evictable suffixes
        \State Freed memory $M_{\mathrm{freed}} \gets 0$
        \While{$M_{\mathrm{freed}} < M$ and $Q \neq \emptyset$}
            \State Agent trajectory $\tau \gets Q.\Call{PopFront}{}$
            \If{$\tau$ is active or has no evictable suffix}
                \State \textbf{continue}
            \EndIf
            \State $c \gets \Call{SelectTailChunk}{\tau, M-M_{\mathrm{freed}}}$
            \State $M_{\mathrm{freed}} \gets M_{\mathrm{freed}} + \Call{Evict}{c}$
            \If{$\tau$ still has an evictable suffix}
                \State $Q.\Call{PushBack}{\tau}$
            \EndIf
        \EndWhile
    \end{algorithmic}
\end{algorithm}

\paragraph{Handling stale agent trajectories.} Direct round-robin has one limitation: because it rotates uniformly across all idle trajectories, a completed trajectory may retain a large cached prefix for many eviction rounds, wasting memory that could benefit live agents. Our implementation supports two ways to handle this stale state. First, the application can notify the serving backend when an agent completes, allowing the backend to immediately release the trajectory's unshared cache. Second, when such a lifecycle signal is unavailable, \SYS{} combines LRU with round-robin through \emph{recency tiering}. It partitions idle trajectories by their last-access times, processes tiers from least to most recent, and applies round-robin eviction among trajectories within the same tier. The cross-tier LRU ordering prioritizes trajectories that are more likely to be stale, while within-tier round-robin distributes cache loss among similarly recent agents. The evaluation uses completion signals for both \SYS{} and the completion-aware LRU baseline. It does not evaluate recency tiering.

\subsection{Analytical Comparison of LRU and \SYS{}}
\label{sec:rr-analysis}

We next formalize the benefit of fine-grained eviction under a simplified homogeneous workload with no cross-agent prefix sharing. The analysis first identifies when memory pressure leaves room to retain idle prefixes and then compares how LRU and ideal round-robin eviction distribute prefix hits and recomputation across returning agents.

\paragraph{Workload and memory assumptions.} Consider $n$ live trajectories with equal-sized private prefixes and no cross-agent sharing. We normalize each complete prefix to one unit and hold its size fixed during reclamation. Each agent spends a fraction $s$ of its lifetime idle and a fraction $a=1-s$ active, with $0<s<1$. If agent phases are sufficiently desynchronized, approximately $n_a=na$ trajectories are active and $n_i=ns$ are idle at a given time. Let $m$ be the cache capacity in prefix-equivalents. Eviction is unnecessary when $m\geq n$. When $m\leq na$, active demand consumes all capacity or exceeds it, leaving no space for idle prefixes and creating a risk of thrashing~\citep{chen2026concur,kang2026thunderagent}. Distributing eviction across idle agents is therefore useful only when
\begin{equation}
    na < m < n.
    \label{eq:rr-regime}
\end{equation}
Here, $na$ is a fluid approximation. A deployment must also account for bursts in the instantaneous active set, which this mean-based model does not capture.

Within this operating regime, the remaining capacity determines how much of each idle prefix can be preserved. After pinning the $na$ active trajectories, the system can retain $m-na$ prefix-equivalents for the $ns$ idle trajectories. Ignoring chunk-size effects, ideal round-robin eviction retains the same fraction of every idle prefix:
\begin{equation}
    h = \frac{m-na}{ns},
    \qquad
    q = 1-h = \frac{n-m}{ns},
    \label{eq:rr-hit-miss}
\end{equation}
where $h$ is the retained fraction and $q$ is the fraction of the old prefix that must be recomputed on return. For comparison, we model LRU's concentrated eviction as retaining $m-na$ idle prefixes completely and removing the remaining $n-m$, ignoring a possible partially retained boundary prefix. For a uniformly selected returning idle agent, the normalized old-prefix recomputation cost $X_\pi$ under policy $\pi\in\{\mathrm{LRU},\mathrm{RR}\}$ is then
\begin{equation}
    X_{\mathrm{LRU}} =
    \begin{cases}
        1, & \text{with probability } q,\\
        0, & \text{with probability } 1-q,
    \end{cases}
    \qquad
    X_{\mathrm{RR}} = q.
    \label{eq:rr-cost-distribution}
\end{equation}

Both policies have the same mean recomputation cost, $\mathbb{E}[X_\pi]=q$, but distribute it differently. LRU places the full cost on a fraction $q$ of returning agents, whereas ideal round-robin spreads it equally. If prefill latency is approximately proportional to uncached prefix length, this reduces the worst-case old-prefix recomputation component of TTFT from $1$ to $q$. At percentile $p$, LRU has a full miss when $q>1-p$, while round-robin incurs a $q$-fraction miss. When $q\leq1-p$, LRU's $p$-quantile is instead zero, so round-robin does not improve every quantile. This comparison excludes new prompt tokens, queueing, and other components of measured TTFT.

The useful regime is consequently bounded on both sides. As $m\rightarrow na$, $q\rightarrow1$ and no policy can retain meaningful idle state. As $m\rightarrow n$, full-prefix misses under LRU become too rare to affect a fixed tail percentile. \SYS{} is most useful between these extremes, where active contexts fit but concentrated misses remain frequent enough to dominate the tail.

\FloatBarrier

\section{Evaluation}
\label{sec:experiments}

We implement \SYS{} in SGLang~\citep{sglang} and evaluate it on conversational and coding-agent benchmarks.

\paragraph{Experimental setup.}
We run our experiments on NVIDIA H100 GPUs connected by NVLink. For conversational agents, we serve Qwen3-Coder-30B-A3B-Instruct~\citep{yang2025qwen3} in BF16 with tensor parallelism two (TP2) and run 222 banking task trajectories from the local $\tau^2$-bench suite~\citep{shi2026tau}. For coding agents, we serve Qwen3-8B on SWE-bench~\citep{jimenez2024swebench} under prefill--decode (PD) disaggregation. The 1P3D configuration uses one TP1 prefill worker and three TP1 decode workers, with prefix caching on the prefill worker.

We compare \SYS{} against LRU and a completion-aware LRU variant. LRU is the default prefix-cache eviction policy in serving engines such as vLLM~\citep{vllm} and SGLang~\citep{sglang}. The variant, which we denote as \emph{LRU + Completion Signals}, allows the frontend to notify the serving backend when an agent completes so that the backend can immediately reclaim that agent's unshared cache. \SYS{} uses the same completion signals together with 1,024-token round-robin eviction chunks for live agents.

We vary concurrency to control the number of admitted agents~\citep{kang2026thunderagent,chen2026concur}.
To ensure a fair comparison, we first execute each benchmark once to record its input contexts, output lengths, and inter-turn delays. We then replay the resulting trace under every policy, preserving the recorded input contexts, output lengths, request order, and inter-turn gaps and thereby reproducing both context growth and the idle periods between agent turns~\citep{pan2026agentreplay}.
The replay initially fills the available agent slots. Once the target concurrency is reached, it admits a new agent only after a resident agent exits. Waiting agents remain admitted while executing tools or awaiting user responses, so their cached histories continue to compete for memory.

We report P99 TTFT and P99 uncached prompt tokens as our two primary metrics. TTFT measures the time from request arrival until the first output token is ready and captures user-visible tail latency. For each request, uncached tokens equal the prompt length minus the cached prefix length and quantify the input requiring prefill after prefix lookup, including new tokens and uncached history. Together, the metrics capture both the outcome of cache eviction and the recomputation that \SYS{} is designed to reduce.

\begin{figure}[!t]
    \centering
    \includegraphics[width=\linewidth]{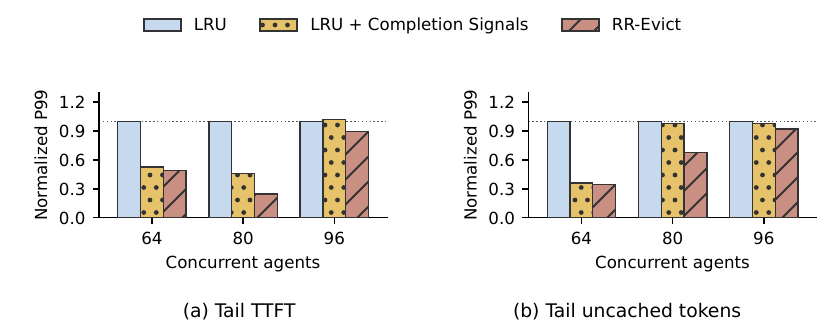}
    \caption{Tail TTFT and uncached prompt tokens on $\tau^2$-bench under different concurrency levels. Results for LRU, LRU + Completion Signals, and \SYS{} are normalized to LRU at the same concurrency. Lower is better.}
    \label{fig:tau-concurrency}
\end{figure}

\paragraph{Results on conversational benchmarks.}
Conversational agents frequently pause for user feedback, leaving many admitted trajectories idle and creating the sparse activity analyzed in Section~\ref{sec:rr-analysis}. Figure~\ref{fig:tau-concurrency} compares LRU, LRU + Completion Signals, and \SYS{} on this workload. Across all evaluated concurrency levels, \SYS{} consistently outperforms both baselines. Relative to LRU, it reduces P99 TTFT and P99 uncached tokens by up to 75.4\% and 65.7\%, respectively. Relative to LRU + Completion Signals, the corresponding reductions are up to 46.9\% and 30.6\%. At lower concurrency, the incremental benefit over completion signals is smaller, consistent with completed-state reclamation relieving much of the memory pressure. At the highest concurrency, the narrower benefit is consistent with active demand leaving less capacity for idle prefixes.

\begin{figure}[!t]
    \centering
    \includegraphics[width=\linewidth]{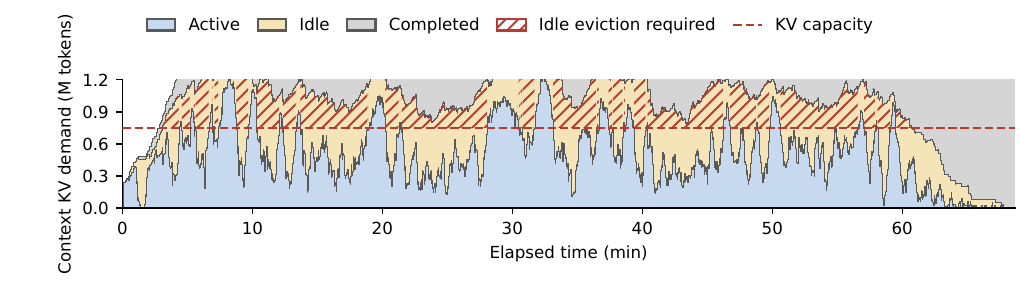}
    \caption{Token-derived logical context-KV demand on $\tau^2$-bench at concurrency 80. The stack separates active requests, live idle trajectories, and completed trajectories. The dashed line marks KV-cache capacity. Red hatching marks intervals in which active KV fits but active plus idle KV does not, so a portion of live idle state must be evicted.}
    \label{fig:kv-state-timeline}
\end{figure}

\paragraph{Trace-level memory pressure.}
Figure~\ref{fig:kv-state-timeline} further explains the $\tau^2$-bench results from the concurrency-80 run. We derive each trajectory's logical KV demand from its prompt tokens and classify it as active while a request is in flight, idle between turns, and completed after its final matched turn. Completion signals allow the system to release completed trajectories directly. The trace nevertheless contains sustained intervals in which active demand remains below capacity but active plus live idle demand exceeds it, as indicated by the hatched region. During these intervals, active requests fit, yet the system must evict part of the live idle state. This is precisely the regime in Equation~\ref{eq:rr-regime}: eviction is necessary, but enough capacity remains for \SYS{} to spread the loss across idle trajectories instead of fully evicting a few of them.

\begin{figure}[!t]
    \centering
    \includegraphics[width=\linewidth]{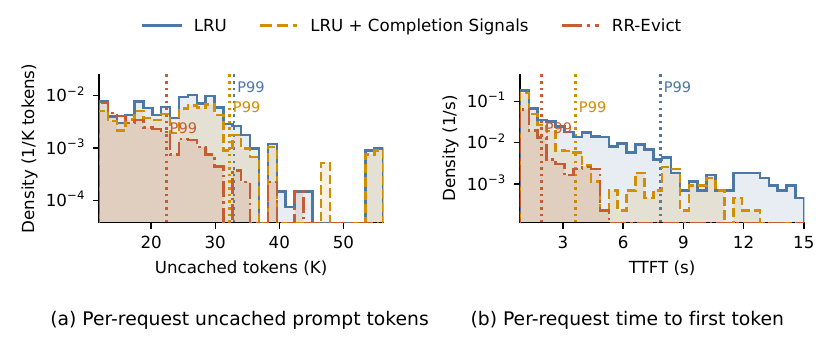}
    \caption{Upper-tail probability densities for per-request uncached prompt tokens and time to first token on $\tau^2$-bench at concurrency 80. The panels compare LRU, LRU + Completion Signals, and \SYS{} using shared bins and full-sample normalization. Dotted lines mark P99.}
    \label{fig:tail-distributions}
\end{figure}

\paragraph{Uncached token and TTFT distributions.}
Figure~\ref{fig:tail-distributions} shows the tail uncached token and TTFT distributions of $\tau^2$-bench requests at concurrency 80.
\SYS{} substantially shortens the extreme uncached-token tail, which in turn shifts the TTFT tail to lower values. The TTFT improvement is more pronounced than the reduction in uncached tokens. This larger shift is consistent with \SYS{} avoiding exceptionally long prefills that would otherwise introduce head-of-line blocking and additional queueing delay for subsequent requests. Some long cache-miss events remain even under \SYS{}. The active-demand bursts in Figure~\ref{fig:kv-state-timeline} suggest a possible explanation: less capacity remains for idle histories during these intervals, so returning agents may still incur large prefills.

\begin{figure}[!t]
    \centering
    \includegraphics[width=\linewidth]{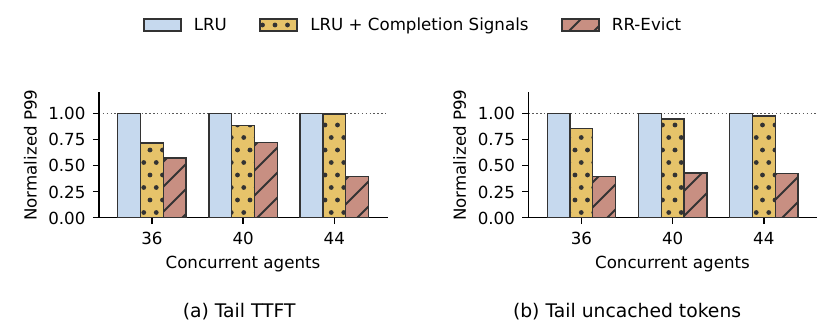}
    \caption{Tail TTFT and uncached prompt tokens for Qwen3-8B on SWE-bench under 1P3D disaggregation at different concurrency levels. Results for LRU, LRU + Completion Signals, and \SYS{} are normalized to LRU at the same concurrency. TTFT includes prefill queueing. Lower is better.}
    \label{fig:pd-coding}
\end{figure}

\paragraph{Results on PD-disaggregated coding agents.}
Coding agents typically spend less time in tool calls~\citep{li2025continuum}, yielding less inter-turn sparsity for \SYS{} to exploit directly. PD disaggregation creates an additional idle interval for prefix caching: reusable prefix KV remains on the prefill worker while decoding proceeds elsewhere. Figure~\ref{fig:pd-coding} reports results on SWE-bench, where \SYS{} consistently outperforms both baselines across all evaluated concurrency levels. Relative to LRU, it reduces P99 TTFT and P99 uncached tokens by up to 60.7\% and 60.8\%, respectively. Relative to LRU + Completion Signals, the reductions reach 60.5\% and 56.2\%. Completion signals alone have little effect at the highest concurrency, while distributing eviction across live agents reduces both cache misses and prefill queueing. As a result, the TTFT benefit becomes especially pronounced under higher load.

\section{Related Work}
\label{sec:related-work}

\paragraph{LLM serving systems.} LLM serving systems optimize scheduling, operator execution, and KV-cache management to reduce latency and improve hardware utilization. At the scheduling layer, Orca~\citep{yu2022orca} enables continuous batching through iteration-level scheduling. Chunked-prefill techniques~\citep{agrawal2024taming,holmes2024deepspeed} split long prefills into chunks and mix the resulting prefill chunks with decoding, reducing interference between the two phases. Prefill-decode disaggregation~\citep{zhong2024distserve,patel2024splitwise} places the two phases on separate instances to further eliminate such interference.
At the operator-execution layer, prior work optimizes GPU kernels and compilers~\citep{dao2022flashattention,dong2024flex,pan2025fasttree,kamath2025pod,sanovar2025leanattention,wu2025mirage,cheng2025mirage,zheng2023bladedisc,xia2023flash,ye2026cake}, with FlashInfer~\citep{ye2025flashinfer} as a representative serving-oriented kernel library.
For KV caches, vLLM~\citep{vllm} reduces fragmentation with PagedAttention, and SGLang~\citep{sglang} captures prefix sharing with a radix tree. CachedAttention~\citep{gao2024cost} and Pensieve~\citep{yu2025stateful} preserve KV state across conversation turns, while Mooncake~\citep{qin2025mooncake}, LMCache~\citep{liu2025lmcache}, and SGLang HiCache~\citep{xie2026strata} expand capacity through pooled or hierarchical storage. These systems target general request streams rather than long-running agents, whose private histories repeatedly grow across tool or user interactions.

\paragraph{Agent serving systems.} As agent applications become increasingly important, recent work has tailored LLM serving to their execution structure. Parrot~\citep{lin2024parrot}, Ayo~\citep{tan2025towards}, and HedraRAG~\citep{hu2025hedrarag} model workflows as dependency graphs to enable compile-time optimization. Agentix~\citep{luo2026agentix} proposes program-level scheduling to reduce head-of-line blocking. SmoothAgent~\citep{pan2026smoothagent} and FlashAgents~\citep{fang2026flashagents} exploit agent structure to prepare or overlap prefills.
ThunderAgent~\citep{kang2026thunderagent} and CONCUR~\citep{chen2026concur} prevent overload-induced memory thrashing through admission and concurrency control.
FlashEvolve~\citep{hu2026flashevolve} uses asynchronous stage orchestration to accelerate agent self-evolution.
These systems do not specifically optimize KV-cache management for agentic workloads, which is the focus of this paper.
InferCept~\citep{abhyankar2024infercept} and Continuum~\citep{li2025continuum} use tool-duration estimates to guide cache retention, while KVFlow~\citep{pan2025kvflow} and ScaleSim~\citep{pan2026scalesim} rely on application information like workflow structures. Such signals may be unavailable or inaccurate in practice. In contrast, \SYS{} distributes cache loss across idle agents without workflow graphs or tool-execution-time predictors, preserving partial prefixes for more agents.

\section{Conclusion and Discussion}
\label{sec:conclusion}

Long-horizon agents expose a key limitation of LRU: synchronized accesses concentrate eviction on a few agents, causing large prefix recomputations and tail-TTFT spikes. \SYS{} distributes round-robin tail eviction across idle agents, preserving partial prefixes without predicting future arrivals. Our analysis identifies its useful regime, where active contexts fit but all live histories do not. Across the tested colocated and PD-disaggregated agent workloads, our SGLang implementation reduces P99 TTFT by up to 75.4\% and P99 uncached prompt tokens by up to 65.7\% relative to LRU.

Our evaluation is limited to relatively short idle intervals and prefix caches resident in GPU HBM. Future work will evaluate \SYS{} at larger scale, where agents may wait much longer for user feedback and their idle state may need to spill to CPU memory or SSD, while accounting for KV transfer and reload costs.

\section*{AI Use Statement}

Generative AI tools were used to assist with language editing, document organization, code implementation, and the development of plotting scripts. The authors independently designed the method and experiments, produced the measurements, and verified all claims. The authors take full responsibility for the final manuscript.

\bibliography{references}
\bibliographystyle{vendor/iclr2027/iclr2027_conference}


\end{document}